\documentclass{article}
\usepackage{graphicx} 

\title{\bf Six Levels of Privacy: A Framework \\
for Financial Synthetic Data}
\author{Tucker Balch, Vamsi K. Potluru, Deepak Paramanand, Manuela Veloso}
\date{}

\begin{document}

\maketitle

\centerline{J.P. Morgan Chase \& Company AI Research}

\vspace{0.25in}


\begin{abstract}
Synthetic Data is increasingly important in financial applications.
In addition to the benefits it provides, such as improved financial
modeling and better testing procedures, it poses privacy risks as well. 
Such data may arise from client information,
business information, or other proprietary sources that
must be protected.
Even though the process by which Synthetic Data is generated 
serves to obscure the original data to some degree, 
the extent to which privacy is preserved is hard to assess.
Accordingly, we introduce a hierarchy 
of ``levels''
of privacy that are useful for categorizing
Synthetic Data generation methods and the
progressively improved protections they offer.
While the six levels were devised in the context of financial
applications, they may also be appropriate for other
industries as well.
Our paper includes: A brief overview of Financial Synthetic Data, how
it can be used, how its value can be assessed, privacy risks, and privacy
attacks.  We close with details of the ``Six Levels'' that include defenses
against those attacks.
\end{abstract}

\section{Introduction}

As the name suggests, Synthetic Data is artificially
generated rather than produced by real world events. 
Synthetic Data is created via two primary methods, namely: 1) By 
{\it transforming}
real data, or 2) By {\it simulation} of real processes.
We refer the reader to the  rich literature
on Synthetic Data and the many mechanisms for creating it~\cite{assefa2020generatingb}.  In financial applications we
focus on three key uses for Synthetic Data:

\begin{enumerate}
    \item{\bf Liberate data:} Depending on its source, the
        sensitivity or risk associated with particular types of data
        can be significantly reduced or eliminated when transformed
        to synthetic form.  We might be able to, accordingly, share it more
        freely and with less risk. We refer to this as 
        ``liberating data.''
    \item{\bf Augment for training:} Synthetic Data can be used
        to augment real data used for training
        models in order to fill gaps in the coverage of the
        data.  In some cases the models trained in this way
        perform better than those without augmentation.
    \item{\bf Testing:} With Synthetic Data we have the advantage of being 
        able to control the
        generation so that we know its properties and 
        contents.  If for instance,
        we want to test a fraud detection algorithm, we
        can ``plant'' known fraudulent patterns in the data
        to check if an algorithm flags them.  Synthetic Data
        can also be used to explore the ``corner cases'' 
        to see of the processes that use the data break under stress.
\end{enumerate}

The value of Synthetic Data for each of the above uses may vary
according to the application.  Three different properties of the data contribute to
an assessment of its value.
As you will see, these properties are sometimes confounding: It is usually not possible 
for a dataset to score well along
all three dimensions at once.  The dimensions include:

\begin{itemize}
    \item{\bf Realism:} How realistic is the data, in the sense that it matches
    the real process or dataset that we seek to emulate?  In general, the higher the
    fidelity of the Synthetic Data, the more useful it is for downstream processes,
    but at the cost of reduced privacy.
    \item{\bf Privacy:} How easy is it for an adversary to ``reverse engineer'' the dataset
    to infer properties of the original data?  In some cases, it is possible to
    discover specific private information about individual records in the original data
    even though they are not present in the Synthetic Data (see Section 2: Privacy Attacks).
    Other proprietary or competitive information might also be revealed such as
    the distributional properties of data elements like age, salary, or credit rating of a client list.
    \item{\bf Utility:} How well does the data serve the purpose for which it was created?
    As one example, we might want to use the data to augment real data in the training of
    an ML model.  We would evaluate utility in this case by measuring the uplift
    the data provides for the model: E.g., Are its predictions now more accurate?
    In another case, we might be using the data to test an existing model or process,
    say for processing credit card transactions.  These tests might be aimed at discovering ``breaks''
    in the data processing pipeline (e.g., are large, or negative transactions handled appropriately?)
\end{itemize}

The metrics are interrelated, for instance: Increased realism usually suggests
reduced privacy; Increased privacy may degrade utility.
Note that while one might assume realism is the most important 
factor, this is not always the case.  If we are testing a product
or process and we only use real, or historical data, we might not expose
flaws regarding how the system would respond to new, unexpected scenarios.

In the next section we consider some of the risks regarding privacy for financial data.

\section{Privacy risks for financial data}

Financial institutions are appropriately protective of their
data and the data they hold for their clients. Data sharing between various lines of business within a
company, and potentially, externally with clients or vendors, is governed by
regulations and internal guidelines. These controls are designed
to protect clients’ sensitive information and protect firms from 
the unauthorized sharing of 
MNPI (Material Non-Public Information), as well as litigation, reputation, and competitive risks. 

Here we review some prominent risks and 
relevant regulations that apply to financial institutions.
While specific to this industry, these regulations are
representative of those many businesses face.

\begin{itemize}

\item{\bf Fair Credit Reporting Act (FCRA):}

This U.S. law requires that information collected by consumer reporting agencies (e.g. credit bureaus) cannot be provided to anyone who does not have a purpose specified in the Act. In particular, the data cannot be used for other purposes
even if data that identify an individual are removed.
In addition, the data user must ensure that identity cannot be 
inferred using other non-Personally Identifiable Information (PII) data fields. 

\item{\bf Regulation on Unfair, Deceptive or Abusive Acts or Practices (UDAAP):} In many cases consumers and
clients can specify how their personal data can be used.
Sharing such data
is a UDAAP violation
if used or shared in a manner contrary to the choices made by, or representations made to, consumers or clients. In particular, in many settings data is subject to privacy elections 
made by consumers.

\item{\bf Litigation risks:}  Inappropriate release of data or functions of data (e.g., models trained on data, insights from data, or synthetic data resembling these datasets) that reveal PII  or statistics (e.g., global characteristics) of the data, may pose litigation risks. This is particularly important in the context of data sourced from external vendors: Use of such data is typically constrained by contracts that precisely define the scope of the use.

\item{\bf Competitive risks:} Publishing data that reveals the characteristics of a firm's client base or industries and publicly traded companies the firm has interest in, may pose competitive, antitrust and increased insider trading risks. This might apply even if the published data is synthetic.

\end{itemize}


\section{Privacy attacks}

In order to appropriately assess the protections privacy measures might
provide, we must consider how data might be exploited or "attacked" by an adversary \cite{Sun_2023}. 
We assume there exists an adversary 
who aims to extract  private information from
Synthetic Data or from some other output model output. 
Each type of attack is characterized by
assumptions including: What information is available to 
the adversary? What information should be protected?
What is the goal of the attack? 
Here we enumerate 
the most relevant attacks.  Also, see Table \ref{table: privacy attacks vs reg risks} for an
analysis of attacks versus regulatory risks.

\begin{itemize}

\item{\bf Reconstruction attacks} Also known
as attribute inference attacks.
Reconstruction attacks are characterized by an adversary in possession of partial knowledge of a set of features with the aim to recover \textit{sensitive} features or the full data sample. For example, if some columns matching public information for an 
individual (e.g. from voter registration data) correspond to an entry in the  
candidate dataset that also includes private attributes (e.g. credit card billing records), 
the presence of the individual can reveal the values of the private attributes for that 
person. \cite{narayanan2007break}.

\item{\bf Membership inference attacks (MIAs) }
In many cases the presence of an individual’s data in a dataset by itself can reveal sensitive information. The adversary's task in MIA   is to
infer whether an individual was present in 
the training dataset or not \cite{DBLP:journals/corr/ShokriSS16}. 
An adversary with 
knowledge of an individual’s presence in the dataset can further exploit that knowledge in
linkage (or reconstruction) attacks to identify sensitive attributes of that individual. 
Thus, MIA can be used as a stepping stone to launch other types of attack.

\item{\bf Property inference attacks }
Property inference represents the ability to extract properties of the original dataset from the corresponding synthetic data. 
In general, property inference refers to learning summary statistics of the original data (e.g. mean value, quantiles, histograms etc.) under the assumption of access to Synthetic Data only. Note that preventing property inference attacks necessarilty
degrades fidelity of the synthetic data \cite{lin2023summary}.

\end{itemize}

\begin{table}[]
{\footnotesize
\begin{tabular}{c|c|c|c|c|}
\cline{2-5}
& FCRA       & UDAAP      & Litigation Risk & Competitive Risk \\ \hline
\multicolumn{1}{|c|}{Membership Inference Attack} & Applicable & Applicable & Applicable      & N/A              \\ \hline
\multicolumn{1}{|c|}{Attribute Inference Attack}  & Applicable & Applicable & Applicable      & N/A              \\ \hline
\multicolumn{1}{|c|}{Property Inference Attack}   & N/A        & N/A        & Applicable      & Applicable       \\ \hline
\multicolumn{1}{|c|}{Model Inference Attack}      & Applicable        & Applicable          & Applicable               & Applicable                \\ \hline
\end{tabular}
\caption{Privacy attacks on synthetic data can lead to breach of various regulations in financial
applications.}\label{table: privacy attacks vs reg risks}
}
\end{table}

\newpage

\section{Privacy levels}

Now we introduce a six-level
privacy defense hierarchy and discuss the privacy attacks, utility implications, and potential privacy guarantees for each level. Each level corresponds to a group of defense mechanisms with increasingly stronger privacy protections.

These levels can provide guidance to businesses regarding
the security and utility of their Synthetic Data.
For instance, they might choose to allow internal sharing
of Level 2 data if it arises from a non-critical source,
but require Level 4 protections for more sensitive data. 
The relevant privacy level should be determined according to
the use case to balance multiple objectives
such as the business goal, security, speed of generation, and utility.

In the first 4 levels, we consider methods where the data is {\it transformed} from
the original dataset to the Synthetic Data.  In the figures, the original data appears
on the left, and the arrows indicate how the data is transformed.  We focus on tabular
data in these examples, but the principles can apply to other types of
data.

\begin{figure}[hbt!]
\center{
\includegraphics[width=0.9\linewidth]{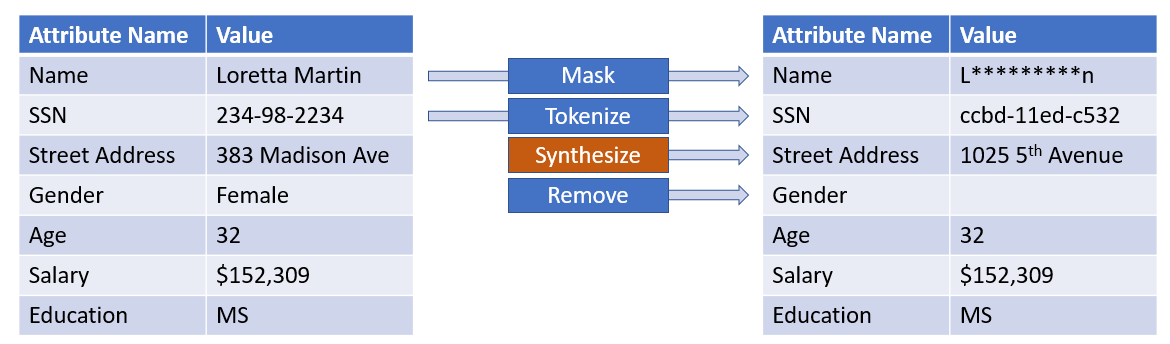}
}
\caption{Privacy Level 1: Obscure PII}
\end{figure}

\subsection{Privacy Level 1: Obscure PII}  
Examples of mechanisms at this level include dropping, replacing, masking, or anonymizing the PII attributes. Since this approach does not modify non-PII attributes in any way, it dones not reduce the utility of downstream tasks and accordingly
there is no utility degradation. This however represents weak privacy protection as data remains vulnerable to reconstruction attacks \cite{narayanan2007break}.

\newpage

\begin{figure}[hbt!]
\center{
\includegraphics[width=0.9\linewidth]{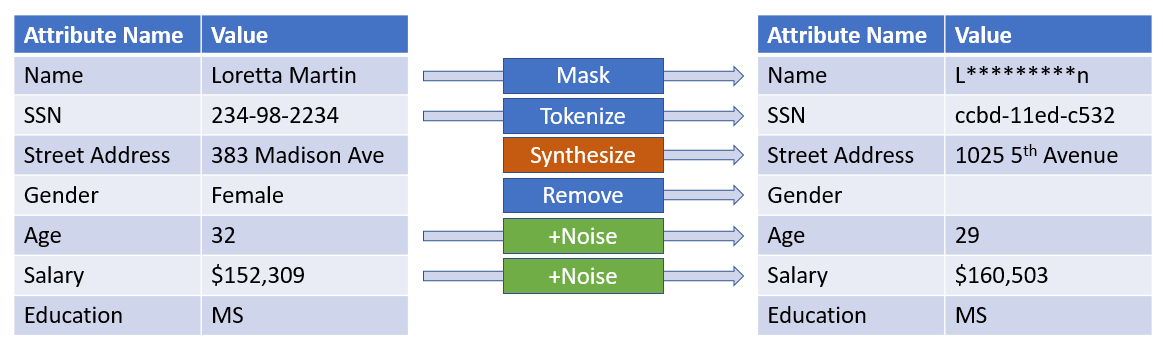}
}
\caption{Privacy Level 2: Obscure PII + noise}
\end{figure}

\subsection{Privacy Level 2: Obscure PII + noise} 
In addition to obscuring PII columns, we can deliberately add
noise to other attributes to reduce the effectiveness of potential attacks. 
Differential privacy techniques, for instance, can provide formal guarantees against MIA.

Another approach involves randomly ``swapping''
data between entries.  So for instance, in a demographic dataset, the
ages of the included individuals might be reordered randomly in the
records.  This technique aims to provide plausible but randomized data
by making it more difficult for an adversary to infer any information regarding 
any particular individual. These techniques
aim to elevate privacy while preserving the utility of the data to a downstream 
task. 

Depending on the amount of noise and the downstream task, some degree of utility degradation is expected. 

\newpage
\begin{figure}[hbt!]
\center{
\includegraphics[width=0.9\linewidth]{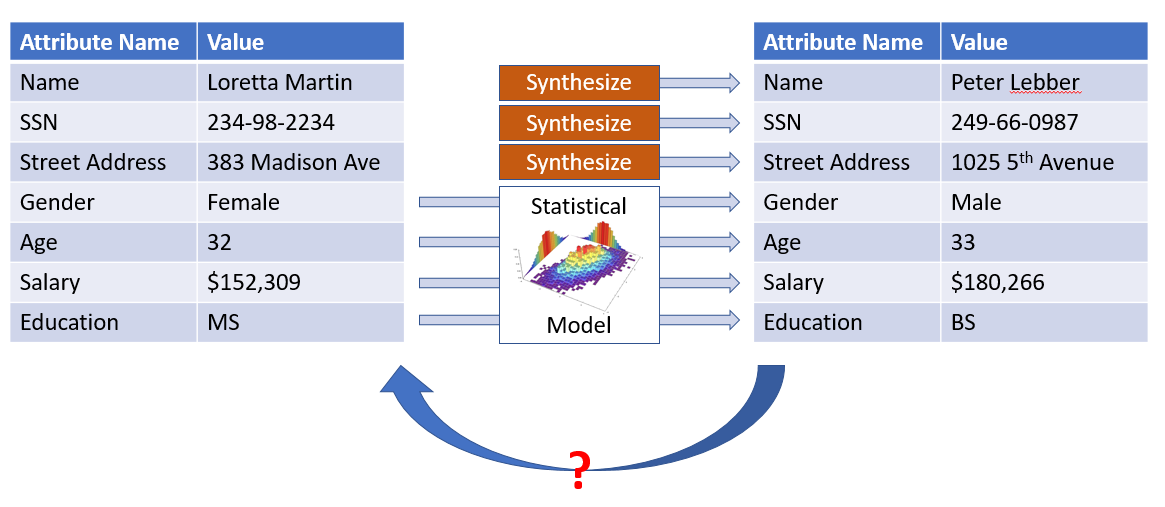}
}
\caption{Privacy Level 3: Generative modeling. The question mark suggests the possibility of
reverse-engineering the data.}
\end{figure}

\subsection{Privacy Level 3: Generative modeling} 
Note that Privacy Levels 1 and 2 involve row-by-row transcription of the original data
(with obfuscation or noise as appropriate).
Accordingly, such datasets cannot be larger than the original.

With Level 3 we move to {\it generative} techniques where we analyze the original
data to build a model that can create new data.
Example approaches include Gaussian copula, and Generative-Adversarial-Networks (GAN)
\cite{7796926, goodfellow2014generative, Park_2018}. Other
methods use differential privacy techniques to offer additional guarantees
\cite{asghar2019differentially, xie2018differentially, yoon2018pategan}.
In our own work, we have
introduced a KD-tree-based formulation
to model the data that offers additional protections as well \cite{KDTree2023differentially}.

All these methods enable the creation of new data elements
distinct from the original data. They offer stronger protection
than in Level 1 or Level 2, but are still potentially
subject to attack.  The risk is
increased when the relative size of the generated data to 
the original data is large: For example, if we generate one million 
samples using an original dataset
of only 1,000 we would expect to see generated samples clustering around the
samples in the original data.


\newpage
\begin{figure}[hbt!]
\center{
\includegraphics[width=0.9\linewidth]{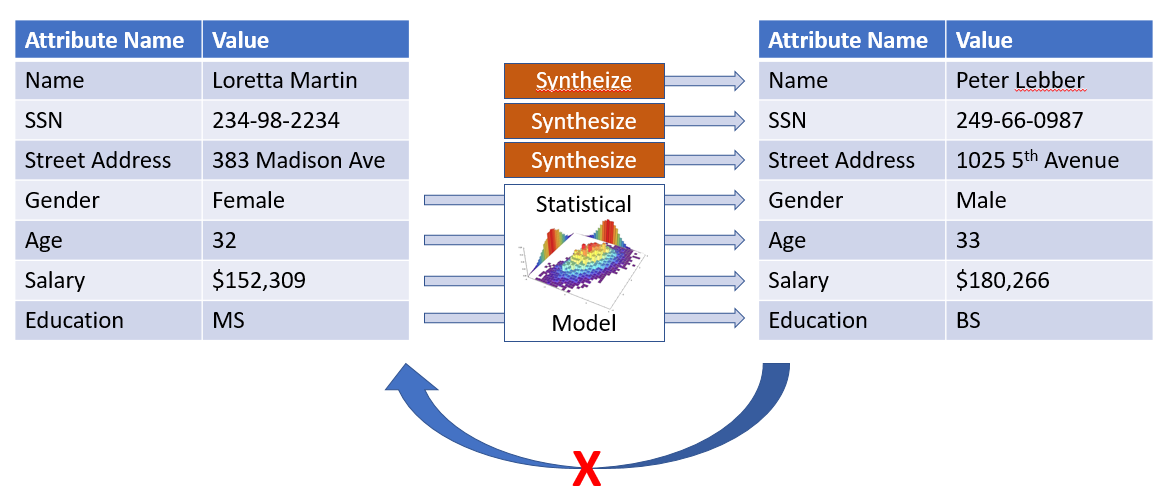}
}
\caption{Privacy Level 4: Generative modeling + testing}
\end{figure}

\subsection{Level 4: Generative modeling + testing} 

For Level 4 we add explicit testing of each generated dataset to validate its
resistance to specific attacks.
The particular tests  and the corresponding scores
required to ``pass'' depend on the data and the application.  For instance,
it may be acceptable for certain properties of the data to ``leak'' while others should not.
To operationalize this, we leverage published attack algorithms, then score the 
data depending on the success of the attack.

While it is hard to specify which test and which score would be necessary
to achieve Level 4 privacy in all cases, the important and
critical difference above Level 3, is the fact that the data is explicitly tested.
The test and the scoring criteria must be determined by the individual business for the
use case. Example scoring criteria measure resistance to membership inference, attribute 
reconstruction, and property attacks. among others~\cite{anonymeter,houssiau2022framework,houssiau2022tapas,belgodere2023auditing,du2024towards}. 

\newpage
\begin{figure}
\center{
\includegraphics[width=0.9\linewidth]{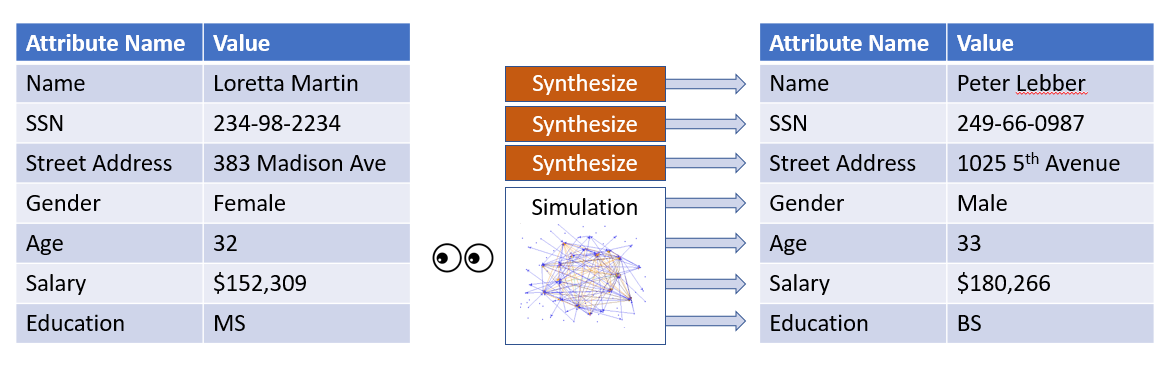}
}
\caption{Privacy Level 5: Calibrated simulation}
\end{figure}
\subsection{Level 5: Calibrated simulation}
In this approach the generation method 
is not trained on real data. In fact, there 
is (usually)
no learning in this approach. Instead, we rely on
simulations governed by rules or knowledge
of the process
that would otherwise generate real data. 
These rules, however, are calibrated with reference to the real process
such that the generated data follows 
some statistical properties of the original, real system. 
As an example, we might use a simulation of the stock market to generate
stock price data.
In our own work, we have developed calibrated simulations of equity markets that correspond to 
Level 5 privacy \cite{vyetrenko2019real}.

Utility degradation depends on the downstream task and the simulation framework. This approach generally represents a strong defense against adversarial attacks.
However, they may be exposed to Property Inference Attacks, because
the simulator is calibrated with respect to statistical properties of 
the real system.

\newpage
\begin{figure}
\center{
\includegraphics[width=0.9\linewidth]{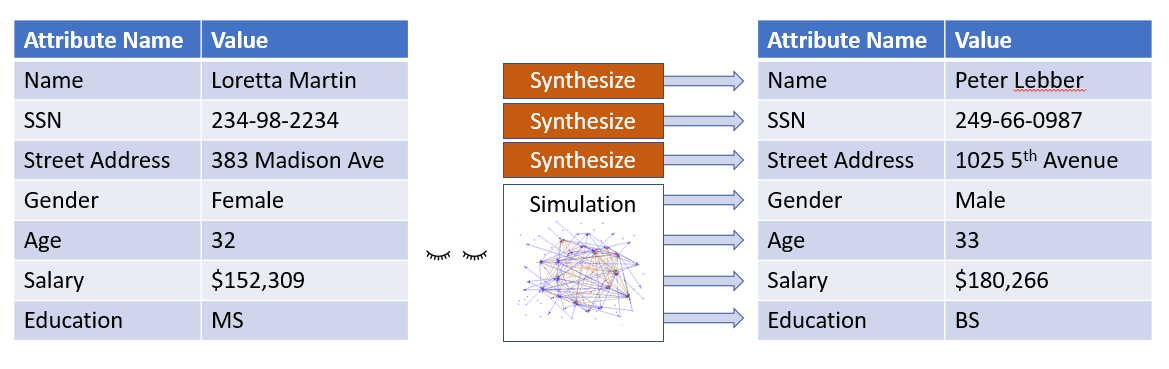}
}
\caption{Privacy Level 6: Uncalibrated simulation}
\end{figure}
\subsection{Level 6: Uncalibrated simulation} 
In this case we may not be aware of the statistical properties
of the modeled system, or we might deliberately avoid adjusting the simulation
to correspond to the properties of the original system.
Even though such a simulation may not provide high fidelity data, it can
still prove quite useful.

An important
use is testing, in which we might use a simulation to
explore all the potential
values of data fields to see if they ``break'' our downstream processes.
Additionally, we might choose to embed known examples
of situations we want to be sure our systems detect (e.g., fraudlent transactions).
Another use is to create what-if scenarios
where we hypothesize the impact of one factor on another, to see if
visualization techniques might enable us to discover those relationships in practice.

In general, this method yields a strong privacy guarantee.  
It remediates one of the consequences of level 5 generation of defence against PIA attacks, 
given that the statistical properties of the data is uncalibrated to the real dataset. 

\section{Summary}

We describe six categories, or levels, of privacy
protection for Financial Synthetic Data provided
by different generation techniques.  The strength of privacy protection 
relates to the resistance the
 technique offers against privacy attacks.  
Such attacks might enable an adversary to infer information about
individual data points in the original data used to train a generator.

The six levels progress
from least secure (Level 1) to most secure (Level 6),  Level 1 depends on
simple masking and obfuscation
(which offers very little protection), while 
Level 6, uncalibrated simulation, provides the 
strongest protection.
We focus specifically on financial data, but
the categorizations may be useful in other industries
(e.g. healthcare)
and generation techniques as well.

\section{Acknowledgements}

We thank Mohsen Ghassemi and Eleonora Krea{\v{c}}i{\'c} for many helpful discussions.

\section{Disclaimer}

This paper was prepared for informational purposes  by the CDAO group of JPMorgan Chase \& Co and its affiliates (“J.P. Morgan”) and is not a product of the Research Department of J.P. Morgan.  J.P. Morgan makes no representation and warranty whatsoever and disclaims all liability, for the completeness, accuracy or reliability of the information contained herein.  This document is not intended as investment research or investment advice, or a recommendation, offer or solicitation for the purchase or sale of any security, financial instrument, financial product or service, or to be used in any way for evaluating the merits of participating in any transaction, and shall not constitute a solicitation under any jurisdiction or to any person, if such solicitation under such jurisdiction or to such person would be unlawful.\\

\center{© 2024 JPMorgan Chase \& Co. All rights reserved.}

\newpage

\bibliographystyle{alpha}
\bibliography{sample,privacy}
\end{document}